\documentclass[conference]{IEEEtran}

\IEEEoverridecommandlockouts

\usepackage{amsmath,amsthm,relsize}
\usepackage{amsfonts, amssymb, cuted}
\usepackage{cleveref}
\usepackage{mathtools}
\usepackage[boxed]{algorithm}
\usepackage[noend]{algpseudocode}
\usepackage{cite}
\usepackage{xcolor}
\usepackage{soul}
\usepackage{graphicx}
\usepackage{tikz}
\usepackage{pgfplotstable}
\usepackage{pgfplots}
\usepackage{nicefrac}
\usepackage{enumitem}
\usepackage{support-caption}
\usepackage{subcaption}
\usepackage[font=footnotesize]{caption}
\usepackage{multirow}
\pgfplotsset{compat=1.15}
\usepackage{relsize}
\usetikzlibrary{patterns}
\usepackage{acro}
\usepackage{color,soul}

\newcommand{\vbar}{\raisebox{.17ex}{\rule{.04em}{1.35ex}}}
\newcommand{\vbarind}{\raisebox{.01ex}{\rule{.04em}{1.1ex}}}
\newcommand{\R}{\ifmmode{\rm I}\hspace{-.2em}{\rm R} \else ${\rm I}\hspace{-.2em}{\rm R}$ \fi}
\newcommand{\T}{\ifmmode{\rm I}\hspace{-.2em}{\rm T} \else ${\rm I}\hspace{-.2em}{\rm T}$ \fi}
\newcommand{\N}{\ifmmode{\rm I}\hspace{-.2em}{\rm N} \else \mbox{${\rm I}\hspace{-.2em}{\rm N}$} \fi}
\newcommand{\B}{\ifmmode{\rm I}\hspace{-.2em}{\rm B} \else \mbox{${\rm I}\hspace{-.2em}{\rm B}$} \fi}
\newcommand{\Hil}{\ifmmode{\rm I}\hspace{-.2em}{\rm H} \else \mbox{${\rm I}\hspace{-.2em}{\rm H}$} \fi}
\newcommand{\C}{\ifmmode\hspace{.2em}\vbar\hspace{-.31em}{\rm C} \else \mbox{$\hspace{.2em}\vbar\hspace{-.31em}{\rm C}$} \fi}
\newcommand{\Cind}{\ifmmode\hspace{.2em}\vbarind\hspace{-.25em}{\rm C} \else \mbox{$\hspace{.2em}\vbarind\hspace{-.25em}{\rm C}$} \fi}
\newcommand{\Q}{\ifmmode\hspace{.2em}\vbar\hspace{-.31em}{\rm Q} \else \mbox{$\hspace{.2em}\vbar\hspace{-.31em}{\rm Q}$} \fi}
\newcommand{\Z}{\ifmmode{\rm Z}\hspace{-.28em}{\rm Z} \else ${\rm Z}\hspace{-.28em}{\rm Z}$ \fi}

\DeclareAcronym{AWGN}{
    short = AWGN,
    long = additive white Gaussian noise,
    list = Additive White Gaussian Noise,
    tag = abbrev
}

\DeclareAcronym{ADMM}{
    short = ADMM,
    long = alternating direction method of multipliers,
    list = Alternating Direction Method of Multipliers,
    tag = abbrev
}

\DeclareAcronym{MGMC}{
    short = MGMC,
    long = multi-group multi-casting,
    list = multi-group multi-casting,
    tag = abbrev
}

\DeclareAcronym{SGMC}{
    short = SGMC,
    long = single-group multi-casting,
    list = single-group multi-casting,
    tag = abbrev
}
\DeclareAcronym{AoA}{
    short = AoA,
    long = angle-of-arrival,
    list = Angle-of-Arrival,
    tag = abbrev
}

\DeclareAcronym{AoD}{
    short = AoD,
    long = angle-of-departure,
    list = Angle-of-Departure,
    tag = abbrev
}

\DeclareAcronym{KKT}{
    short = KKT,
    long = Karush-Kuhn-Tucker,
    list = Karush-Kuhn-Tucker,
    tag = abbrev
}

\DeclareAcronym{MMF}{
    short = MMF,
    long = max-min-fairness,
    list = max-min-fairness,
    tag = abbrev
}

\DeclareAcronym{WMMF}{
    short = WMMF,
    long = weighted max-min-fairness,
    list = max-min-fairness,
    tag = abbrev
}

\DeclareAcronym{BB}{
    short = BB,
    long = base band,
    list = Base Band,
    tag = abbrev
}

\DeclareAcronym{BC}{
    short = BC,
    long = broadcast channel,
    list = Broadcast Channel,
    tag = abbrev
}

\DeclareAcronym{BS}{
    short = BS,
    long = base station,
    list = Base Station,
    tag = abbrev
}

\DeclareAcronym{BR}{
    short = BR,
    long = best response,
    list = Best Response, 
    tag = abbrev
}

\DeclareAcronym{CB}{
    short = CB,
    long = coordinated beamforming,
    list = Coordinated Beamforming,
    tag = abbrev
}

\DeclareAcronym{CC}{
    short = CC,
    long = coded caching,
    list = Coded Caching,
    tag = abbrev
}

\DeclareAcronym{CE}{
    short = CE,
    long = channel estimation,
    list = Channel Estimation,
    tag = abbrev
}

\DeclareAcronym{CoMP}{
    short = CoMP,
    long = coordinated multi-point transmission,
    list = Coordinated Multi-Point Transmission,
    tag = abbrev
}

\DeclareAcronym{CRAN}{
    short = C-RAN,
    long = cloud radio access network,
    list = Cloud Radio Access Network,
    tag = abbrev
}

\DeclareAcronym{CSE}{
    short = CSE,
    long = channel specific estimation,
    list = Channel Specific Estimation,
    tag = abbrev
}

\DeclareAcronym{CSI}{
    short = CSI,
    long = channel state information,
    list = Channel State Information,
    tag = abbrev
}

\DeclareAcronym{CSIT}{
    short = CSIT,
    long = channel state information at the transmitter,
    list = Channel State Information at the Transmitter,
    tag = abbrev
}

\DeclareAcronym{CU}{
    short = CU,
    long = central unit,
    list = Central Unit,
    tag = abbrev
}

\DeclareAcronym{D2D}{
    short = D2D,
    long = device-to-device,
    list = Device-to-Device,
    tag = abbrev
}

\DeclareAcronym{DE-ADMM}{
    short = DE-ADMM,
    long = direct estimation with alternating direction method of multipliers,
    list = Direct Estimation with Alternating Direction Method of Multipliers,
    tag = abbrev
}

\DeclareAcronym{DE-BR}{
    short = DE-BR,
    long = direct estimation with best response,
    list = Direct Estimation with Best Response,
    tag = abbrev
}

\DeclareAcronym{DE-SG}{
    short = DE-SG,
    long = direct estimation with stochastic gradient,
    list = Direct Estimation with Stochastic Gradient,
    tag = abbrev
}

\DeclareAcronym{DFT}{
	short = DFT,
	long = discrete fourier transform,
	list = Discrete Fourier Transform,
	tag = abbrev
}

\DeclareAcronym{DoF}{
    short = DoF,
    long = degrees of freedom,
    list = Degrees of Freedom,
    tag = abbrev
}

\DeclareAcronym{DL}{
    short = DL,
    long = downlink,
    list = Downlink,
    tag = abbrev
}

\DeclareAcronym{GD}{
	short = GD, 
	long = gradient descent,
	list = Gradeitn Descent,
	tag = abbrev
}

\DeclareAcronym{IBC}{
    short = IBC,
    long = interfering broadcast channel,
    list = Interfering Broadcast Channel,
    tag = abbrev
}

\DeclareAcronym{i.i.d.}{
    short = i.i.d.,
    long = independent and identically distributed,
    list = Independent and Identically Distributed,
    tag = abbrev
}

\DeclareAcronym{JP}{
    short = JP,
    long = joint processing,
    list = Joint Processing,
    tag = abbrev
}

\DeclareAcronym{LOS}{
	short = LOS,
	long = line-of-sight,
	list = Line-of-Sight,
	tag = abbrev
}

\DeclareAcronym{LS}{
    short = LS,
    long = least squares,
    list = Least Squares,
    tag = abbrev
}

\DeclareAcronym{LTE}{
    short = LTE,
    long = Long Term Evolution,
    tag = abbrev
}

\DeclareAcronym{LTE-A}{
    short = LTE-A,
    long = Long Term Evolution Advanced,
    tag = abbrev
}

\DeclareAcronym{MIMO}{
    short = MIMO,
    long = multiple-input multiple-output,
    list = Multiple-Input Multiple-Output,
    tag = abbrev
}

\DeclareAcronym{MISO}{
    short = MISO,
    long = multiple-input single-output,
    list = Multiple-Input Single-Output,
    tag = abbrev
}

\DeclareAcronym{MAC}{
    short = MAC,
    long = multiple access channel,
    list = Multiple Access Channel,
    tag = abbrev
}

\DeclareAcronym{MSE}{
    short = MSE,
    long = mean-squared error,
    list = Mean-Squared Error,
    tag = abbrev
}

\DeclareAcronym{MMSE}{
    short = MMSE,
    long = minimum mean-squared error,
    list = Minimum Mean-Squared Error,
    tag = abbrev
}

\DeclareAcronym{mmWave}{
	short = mmWave,
	long = millimeter wave,
	list = Millimeter Wave,
	tag = abbrev
}

\DeclareAcronym{MU-MIMO}{
    short = MU-MIMO,
    long = multi-user \ac{MIMO},
    list = Multi-User \ac{MIMO},
    tag = abbrev
}

\DeclareAcronym{OTA}{
    short = OTA,
    long = over-the-air,
    list = Over-the-Air,
    tag = abbrev
}

\DeclareAcronym{PSD}{
    short = PSD,
    long = positive semidefinite,
    list = Positive Semidefinite,
    tag = abbrev
}

\DeclareAcronym{QoS}{
	short = QoS,
	long = quality of service,
	list = Quality of Service,
	tag = abbrev
}

\DeclareAcronym{RCP}{
	short = RCP,
	long = remote central processor,
	list = Remote Central Processor,
	tag = abbrev
}

\DeclareAcronym{RRH}{
    short = RRH,
    long = remote radio head,
    list = Remote Radio Head,
    tag = abbrev
}

\DeclareAcronym{RSSI}{
    short = RSSI,
    long = received signal strength indicator,
    list = Received Signal Strength Indicator,
    tag = abbrev
}

\DeclareAcronym{RX}{
	short = RX,
	long = receiver,
	list = Receiver,
	tag = abbrev
}

\DeclareAcronym{SCA}{
    short = SCA,
    long = successive-convex-approximation,
    list = Successive-Convex-Approximation,
    tag = abbrev
}

\DeclareAcronym{SG}{
    short = SG,
    long = stochastic gradient,
    list = Stochastic Gradient,
    tag = abbrev
}

\DeclareAcronym{SIC}{
    short = SIC,
    long = successive interference cancellation,
    list = Successive Interference Cancellation,
    tag = abbrev
}

\DeclareAcronym{SNR}{
    short = SNR,
    long = signal-to-noise-ratio,
    list = Signal-to-Noise Ratio,
    tag = abbrev
}

\DeclareAcronym{SDR}{
    short = SDR,
    long = semi-definite-relaxation,
    list = semi-definite-relaxation,
    tag = abbrev
}

\DeclareAcronym{SINR}{
    short = SINR,
    long = signal-to-interference-plus-noise ratio,
    list = Signal-to-Interference-plus-Noise Ratio,
    tag = abbrev
}

\DeclareAcronym{SOCP}{
	short = SOCP, 
	long = second order cone program,
	list = Second Order Cone Program,
	tag = abbrev
}

\DeclareAcronym{SSE}{
    short = SSE,
    long = stream specific estimation,
    list = Stream Specific Estimation,
    tag = abbrev
}

\DeclareAcronym{SVD}{
	short = SVD,
	long = singular value decomposition,
	list = Singular Value Decomposition,
	tag = abbrev
}

\DeclareAcronym{TDD}{
	short = TDD,
	long = time division duplex,
	list = Time Division Duplex,
	tag = abbrev
}

\DeclareAcronym{TX}{
	short = TX,
	long = transmitter,
	list = Transmitter,
	tag = abbrev
}

\DeclareAcronym{UE}{
    short = UE,
    long = user equipment,
    list = User Equipment,
    tag = abbrev
}

\DeclareAcronym{UL}{
    short = UL,
    long = uplink,
    list = Uplink,
    tag = abbrev
}

\DeclareAcronym{ULA}{
	short = ULA,
	long = uniform linear array,
	list = Uniform Linear Array,
	tag = abbrev
}

\DeclareAcronym{UPA}{
    short = UPA,
    long = uniform planar array,
    list = Uniform Planar Array,
    tag = abbrev
}

\DeclareAcronym{WMMSE}{
    short = WMMSE,
    long = weighted minimum mean-squared error,
    list = Weighted Minimum Mean-Squared Error,
    tag = abbrev
}

\DeclareAcronym{WMSEMin}{
    short = WMSEMin,
    long = weighted sum \ac{MSE} minimization,
    list = Weighted sum \ac{MSE} Minimization,
    tag = abbrev
}

\DeclareAcronym{WBAN}{
	short = WBAN,
	long = wireless body area network,
	list = Wireless Body Area Network,
	tag = abbrev
}

\DeclareAcronym{WSRMax}{
    short = WSRMax,
    long = weighted sum rate maximization,
    list = Weighted Sum Rate Maximization,
    tag = abbrev
}

\newtheorem{exmp}{Example}
\theoremstyle{definition}

\newtheorem{rem}{Remark}

\newcommand{\CL}[0]{{\mathcal{L}}}

\newcommand{\CS}[0]{{\mathcal{S}}}
\newcommand{\CT}[0]{{\mathcal{T}}}
\newcommand{\CU}[0]{{\mathcal{U}}}
\newcommand{\CV}[0]{{\mathcal{V}}}

\newcommand{\Bd}[0]{{\mathbf{d}}}

\newcommand{\Bh}[0]{{\mathbf{h}}}

\newcommand{\Bm}[0]{{\mathbf{m}}}

\newcommand{\Bp}[0]{{\mathbf{p}}}

\newcommand{\Br}[0]{{\mathbf{r}}}

\newcommand{\Bu}[0]{{\mathbf{u}}}
\newcommand{\Bv}[0]{{\mathbf{v}}}

\newcommand{\Sfc}[0]{{\mathsf{c}}}
\newcommand{\Sfd}[0]{{\mathsf{d}}}

\newcommand{\Sff}[0]{{\mathsf{f}}}

\newcommand{\Sfk}[0]{{\mathsf{k}}}
\newcommand{\Sfl}[0]{{\mathsf{l}}}

\newcommand{\Sfn}[0]{{\mathsf{n}}}

\newcommand{\Sft}[0]{{\mathsf{t}}}

\newcommand{\Sfv}[0]{{\mathsf{v}}}

\newcommand{\Sfx}[0]{{\mathsf{x}}}

\newcommand{\SfL}[0]{{\mathsf{L}}}

\usepackage{titlesec}
\titlespacing\section{3pt}{6pt plus 4pt minus 2pt}{6pt plus 2pt minus 2pt}
\titlespacing\subsection{3pt}{4pt plus 4pt minus 2pt}{4pt plus 2pt minus 2pt}
\titlespacing\subsubsection{3pt}{3pt plus 4pt minus 2pt}{0pt plus 2pt minus 3pt}

\title{Optimal Fairness Scheduling for Coded Caching\\in Multi-AP Wireless Local Area Networks}

\begin{document}

\author{\IEEEauthorblockN{Kagan Akcay\IEEEauthorrefmark{1}, MohammadJavad Salehi\IEEEauthorrefmark{2}, and Giuseppe Caire\IEEEauthorrefmark{1}} \\
\IEEEauthorblockA{
    \IEEEauthorrefmark{1} Electrical Engineering and Computer Science Department, Technische Universit\"at Berlin, 10587 Berlin, Germany\\
    \IEEEauthorrefmark{2} Centre for Wireless Communications, University of Oulu, 90570 Oulu, Finland \\
    \textrm{kagan.akcay@tu-berlin.de \quad mohammadjavad.salehi@oulu.fi \quad caire@tu-berlin.de}
    }
}

\maketitle

\begin{abstract}
Coded caching (CC) schemes exploit the cumulative cache memory of the users and simple linear coding to turn unicast traffic (individual file requests) into a multicast transmission.
For the originally proposed $K$-user single-server/single shared link network model, CC yields an $O(K)$ gain with respect to conventional uncoded caching with the same per-user memory. While several information-theoretic optimality results for a variety of problems and carefully crafted network topologies have been proved, the gains and suitability of CC for practical scenarios such as content streaming over existing wireless networks have not yet been fully demonstrated. In this work, we consider CC for on-demand video streaming over WLANs where multiple users are served simultaneously by multiple spatially distributed access points (AP). Users sequentially request video ``chunks". The CC scheme operates above the IP layer, leaving the underlying standard physical layer and MAC layer untouched. The cache placement is completely asynchronous and decentralized, and the users are placed at random over the network coverage area. For such a system, 
we consider the region of achievable long-term average delivery rate (defined as the number of video chunks delivered per unit of time) and study the per-user rate distribution under proportional fairness scheduling. We also consider reduced complexity scheduling strategies and compare them
with standard state-of-the-art techniques such as conventional (uncoded) caching and collision avoidance by allocating APs on different sub-channels (i.e., frequency reuse).
\end{abstract}

\begin{IEEEkeywords}
coded caching,
multi-AP communications,
WLAN,
video streaming, 
scheduling.
\end{IEEEkeywords}

\section{Introduction}

The increasing amount of data traffic, especially driven by multimedia applications, has necessitated the development of new communication techniques. One interesting resource is on-device memory; it is cheap and can be used to proactively store a large part of multimedia content, e.g., for video-on-demand (VoD) applications. As a result, many researchers have considered the efficient use of onboard memory in the context of \emph{caching}. Pioneering works in this regard introduced femtocaching~\cite{shanmugam2013femtocaching} and F-RAN models~\cite{park2016joint}. A major breakthrough happened with the introduction of coded caching (CC)~\cite{maddah2014fundamental}, which showed a speedup factor in the achievable rate, scaling with the cumulative cache size in the network, is possible by multicasting carefully created codewords. Later, many works extended the original CC work of~\cite{maddah2014fundamental}, e.g., for wireless~\cite{tolli2017multi}, multi-server~\cite{shariatpanahi2016multi}, multi-antenna~\cite{shariatpanahi2018physical}, D2D~\cite{ji2015fundamental}, shared-cache~\cite{parrinello2019fundamental}, multi-access~\cite{serbetci2019multi}, and combinatorial~\cite{brunero2022fundamental} networks.


While these works are mathematically elegant and often yield information-theoretic optimality results~\cite{yu2017exact},
the application of CC in realistic scenarios remains problematic. For example, most of the works treating CC on wireless networks assume that the physical layer (e.g., multi-user MIMO precoding) can be jointly designed with the CC scheme~\cite{salehi2020lowcomplexity,lampiris2018adding}.
While this is possible in principle, it is highly unlikely that a new wireless standard's PHY and MAC layers are designed as a function of CC. This is because CC, as a content distribution scheme, is implemented at the ``application layer'' (i.e., it is deployed at the video server and at the video clients running on the user devices), 
while the PHY and MAC layers are defined by completely different standardization bodies and must serve a much wider range of applications and traffic types.  

Recognizing this practical constraint, the work~\cite{mozhgan} considered CC ``over IP'', i.e., above an existing network layer capable of multicast routing. This present work starts from the network model of~\cite{mozhgan}, which models a carrier-sense multiple access (CSMA) wireless local area network (WLAN), such as a WiFi setup, and studies the optimization of the per-user content delivery throughput subject to a fairness criterion. 
%
%
%
%
%
In the considered network, a video server transmits data to multiple spatially distributed users through multiple access points (APs), referred to as ``helpers''. 
A collision-type interference model is used, such that packets are lost if a user 
gets the superposition of concurrent packets from different helpers above a certain interference threshold (see Section~\ref{section:sys_model}). For this network model, the authors of~\cite{mozhgan} considered the optimization of the worst-case delivery time over all user requests. Using the multi-round delivery method in~\cite{caire} (which was later proved by~\cite{parrinello2019fundamental} to be optimal for the shared-cache model), they first used graph coloring to construct a reuse pattern for the helpers and optimized the helper-user allocation to minimize the delivery time. Next, they introduced a heuristic scheme dubbed  `Avalanche' to leverage the interference-free state of the users as helpers finished their multi-round delivery. The helper activation in the `Avalanche' scheme was reminiscent of CSMA, where an AP grabs the channel when it detects no other AP is transmitting above its interference threshold.

The schemes in~\cite{mozhgan} are heuristic and do not achieve any optimality metric. Moreover, the single-request assumption of~\cite{mozhgan} is not appealing for real-world implementation of CC in WLAN. This is because, in practice, CC applied to VoD streaming requires the sequential delivery of a long sequence of ``video chunks'' 
forming the streaming session. Each chunk contains a relatively small segment of a video, and the chunks must be delivered at a rate (slightly) higher than the playback rate at which the video buffer is emptied at the user video player application.  Hence, the most important performance metric is the delivery rate expressed as the long-term time-averaged number of chunks delivered 
per unit of time. 

In this work, we consider sequential requests and present a computational way to determine the region of per-user delivery rates. The fairness scheduling problem is formulated as the maximization of a concave component-wise non-decreasing network utility function over the rate region. 
The choice of the network utility function determines the fairness criterion; we focus on proportional fairness~\cite{fairness} here. 
We also provide two heuristic methods based on restricting the throughput region and a greedy algorithmic approach to achieve a sub-optimal solution with reduced complexity. 
Numerical analyses are used to evaluate the performance of the proposed solutions.


\section{System Model}
\label{section:sys_model}

\begin{figure}[t]
        \centering
        \includegraphics[width = 0.8\columnwidth]{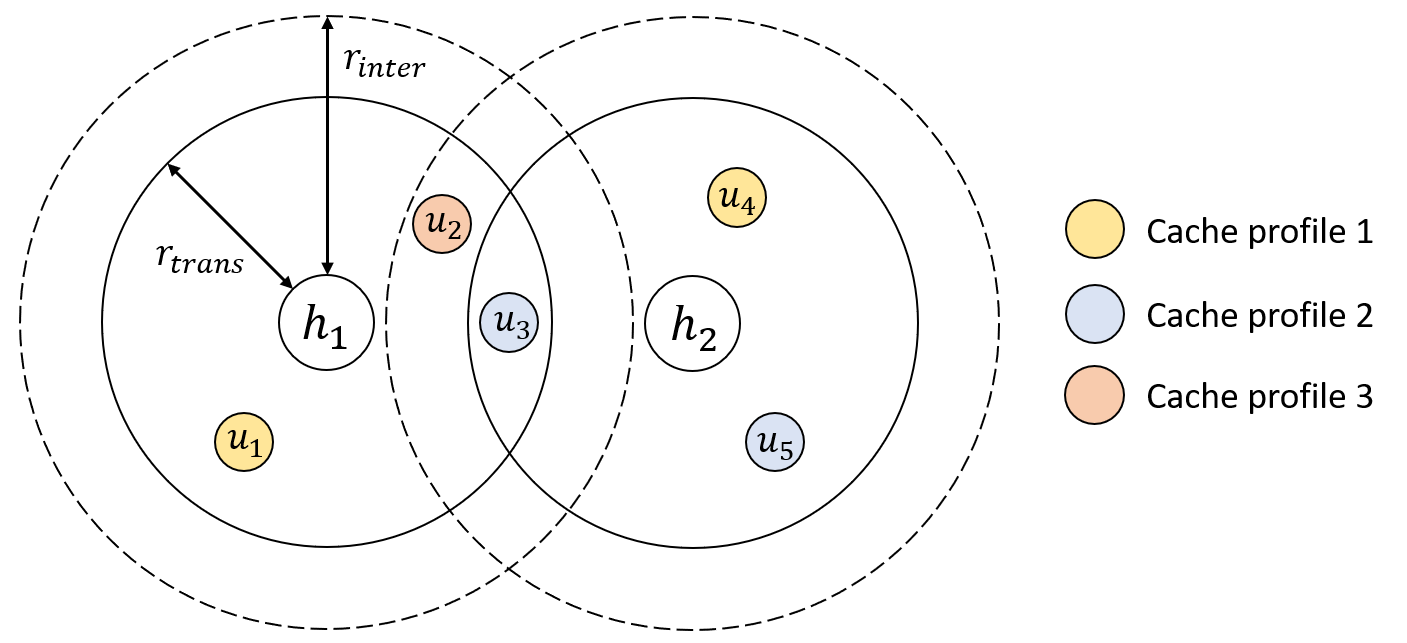} 
        \vspace{-1pt}
    \caption{Example network with $H=2$, $K=5$, $L = 3$, and $t=1$.}
    \label{fig:network_model}
\end{figure}

The system model considered in the paper is similar to the broadcast/collision model of~\cite{mozhgan}: a single server connected to $H$ helper nodes (APs in WLAN) via 
error-free fronthaul links serves the requests of $K$ cache-enabled users. 
We use $h_i$ and $u_k$, $i \in [H]$ and $k \in [K]$, to represent helpers and users, respectively ($[C] \equiv \{1,\cdots,C\}$ for integer $C$). 
Helpers have an effective transmission radius of $r_{\mathrm{trans}}$ and an interference radius of $r_{\mathrm{inter}} \ge r_{\mathrm{trans}}$. 
We call a helper $h_i$ `active' at a given time slot if it is transmitting data in that slot. 
According to the collision model, 
denoting the locations of the helper $h_i$ and user $u_k$ as $\Sfx(h_i), \Sfx(u_k) \in \mathbb{R}^2$, respectively, $u_k$ can successfully decode the transmitted message (packet)
of $h_i$ if and only if 1) $|\Sfx(h_i) - \Sfx(u_k)| \le r_{\mathrm{trans}}$, and 2) there is no other active helper node $h_{i'}$ such that $|\Sfx(h_{i'}) - \Sfx(u_k)| \le r_{\mathrm{inter}}$.
Fig.~\ref{fig:network_model} illustrates a simple example network with $H = 2$ and $K = 5$.

In video streaming, every video file is split into  a sequence of `chunks', each corresponding to a few seconds of video. A streaming user generates a sequence of requests for the corresponding chunks of that video. Each video file is formed by a very large (virtually infinite) number of chunks.\footnote{Imagine movies of duration 1h and 30min, while each video chunk corresponds to 10 seconds of video at the rate of 2Mb/s, i.e., 20Mb of data.} 
The total video library is formed by $N$ files, and 
each user has a cache memory equivalent to $M$ video files. Hence, each user can cache a fraction $\gamma = M/N$ of each chunk of each video. 

System operation consists of two phases, placement and delivery. 
The placement phase is done offline, while the delivery is done during the streaming session. During the placement phase, users' cache memories are filled up with subpackets (i.e., smaller portions) of each chunk. 
In order to allow users to join and leave the network and start their streaming session at any time, it is of fundamental importance that the cache placement is completely decentralized. For this purpose, we use the scheme of \cite{caire,mozhgan} that creates $L$ distinct {\em cache profiles} and lets each user joining the system pick one at random. The number of cache profiles $L$ is an important system design parameter (see later).  
Since the subpacketization is repeated identically for each chunk of each video, we shall simply denote by $W$ a generic chunk. 
Assuming $t = \gamma L$ is an integer,\footnote{If $\gamma L$ is not an integer, the scheme can be modified by cache sharing between two schemes with $t_1 = \lfloor \gamma L \rfloor$, $t_2 = \lceil \gamma L \rceil$. This is well-known in the CC literature and will not be further discussed here due to space limitations.}
every chunk $W$ is partitioned into $\binom{L}{t}$ equal-sized subpackets $W_{\CS}$, 
indexed by all possible subsets $\CS \subseteq [L]$ of size $|\CS| = t$. 
Let $\SfL(u_k) \in [L]$ denote the cache profile picked by user $u_k$ (at random with uniform probability and independently of the other users). If  $\SfL(u_k) = l$, 
user $u_k$ stores every subpacket $W_{\CS}$ for all $\CS \ni l$ (and this is repeated for all chunks of all video files).

\begin{exmp}
\label{exmp:placement}
Consider the network in Fig.~\ref{fig:network_model}. Each chunk of each video file is split into $\binom{3}{1} = 3$ subpackets, and: 1) users $u_1,u_4$ have cache profile~1, and store the first subpacket $W_1$ of each chunk, 2) users $u_3,u_5$ have cache profile~2 and store the second subpacket $W_2$ of each chunk, and 3) user $u_2$ has cache profile~3 and stores the third subpacket $W_3$ of each chunk.
\end{exmp}


During the delivery phase, users reveal their requested video chunks (corresponding to the videos they want to stream). We assume users can start streaming at arbitrary times, and when they start streaming, they repeatedly generate requests for consecutive video chunks. In other words, after receiving every single chunk, each user immediately generates a new request for the subsequent chunk. With this model, we can safely assume that all requested chunks are distinct. In fact, even if two users are streaming the same video, the probability that they start streaming exactly at the same time vanishes with the assumption of very long files. Given the set of user requests, 
the server creates a number of codewords and transmits each one to a subset of users through a selection of active helpers.

\begin{exmp}
\label{exmp:simple_codeword}
    Consider the network in Example~\ref{exmp:placement} and suppose the requests of users $u_1, u_2, u_3, u_4,u_5$ are (the current chunks of) files $A, B, C, D, E$. Users $u_1,u_2,u_3$ can be served by helper $h_1$ provided that helper $h_2$ stays silent. In that case, user $u_1$ needs subpackets $A_2,A_3$, user $u_2$ needs subpackets $B_1,B_2$, and user $u_3$ needs $C_1,C_3$. Hence, helper $h_1$ transmits (in multicast) the codewords $A_2 \oplus C_1$,  $A_3 \oplus B_1$, and $B_2 \oplus C_3$, where $\oplus$ denotes the XOR operation over a finite field. Notice that user $u_1$ has $C_1, B_1$ in its cache and therefore can retrieve $A_2, A_3$ by simple XOR-ing of the received packets $A_2 \oplus C_1$,  $A_3 \oplus B_1$, and so can also do users $u_2$ and $u_3$. At the end of this transmission, users $u_1, u_2 ,u_3$ have obtained their requested chunk, while users $u_4,u_5$ have not received anything. Also, the transmission duration is 3 slots (each slot corresponds to transmitting the equivalent of $1/3$ of a chunk). Hence, the achieved delivery rate vector is $(\frac{1}{3}, \frac{1}{3}, \frac{1}{3}, 0, 0)$. 
\end{exmp}

The rate vector in Example~\ref{exmp:simple_codeword} is referred to as the {\em instantaneous rate vector} corresponding to a particular scheduling decision (in this case, the decision of serving users $u_1, u_2, u_3$ from helper $h_1$ leaving users $u_4$ and $u_5$ unserved). 
%

In general, a network with $H$ helpers has $2^H - 1$ helper activation patterns where at least one helper is active. Let us use vectors $\Bp_j \in \{0,1\}^H$, $j \in [2^H-1]$, to denote all 
activation patterns. The activation pattern determines which users can be served by each active helper and whether they can be served by multicasting codewords.
A codeword can be created for every subset of users $\CV$ of size $0 < |\CV| \le t + 1$ 
if all users in $|\CV|$ are served by the same active helper $h_i$ and have distinct cache profiles. Of course, due to the repetition in cache profile assignments and multiple choices for the size of $\CV$, multiple codeword creation possibilities may exist for each helper.

The server may need to choose multiple activation patterns to serve all the requests because it is likely that some users cannot be served for a specific pattern $\Bp_j$ due to interference from active helpers. 
Moreover, as said, we may have multiple choices for codeword creation with every activation pattern. We let $\Sfc(\Bp_j)$ denote the number of such choices for a given activation pattern $\Bp_j$.
A \emph{scheduling decision} $(j,s)$ corresponds to the selection of 
activation pattern $\Bp_j$ and codeword choice $s \in [\Sfc(\Bp_j)]$. 
As illustrated by the above example, each scheduling decision 
$(j,s)$ results in an instantaneous rate vector $\Br(j,s)$, where 
the element $k$, $k \in [K]$, in $\Br(j,s)$ is the number of chunks per slot obtained by user~$u_k$ under decision $(j,s)$. 
A scheduling policy consists of a sequence of scheduling decisions $\{(j_t, s_t) : t = 1, 2, \ldots\}$. The throughput rate vector achieved by a given scheduling policy is given by 
$\bar{\Br} = \lim_{T \rightarrow \infty} \frac{1}{T} \sum_{t = 1}^T \Br(j_t,s_t)$,
if such a limit exists. It is well-known~\cite{georgiadis2006resource}
that the throughput region of the network is given by the convex hull of all instantaneous 
rate vectors, i.e., 
\[ 
{\cal R} = \mathrm{Conv}(\Br(j,s) : j \in [2^H], s \in [\Sfc(\Bp_j)]), 
\]
and that each point in the throughput region can be achieved by some stationary scheduling policy, i.e.,  a policy that at each slot chooses (independently over the time slots) 
decision $(j, s)$ with probability $a(j,s)$.

Clearly, we are interested in scheduling policies that operate the system on the Pareto boundary of ${\cal R}$. However, the choice of the operating point may correspond to different criteria of optimality. Following~\cite{georgiadis2006resource},
we are interested in
optimizing a network utility function of the per-user throughput rates. By choosing as network utility a concave component-wise non-descending function, one can impose a desired criterion of fairness. For example, we may choose any function from the alpha-fairness family~\cite{fairness}: for $\alpha = 1$, we get proportional fairness; for $\alpha = 0$, we get sum-rate (unfair); and for $\alpha \rightarrow \infty$, we get max-min (hard) fairness. Here, for the sake of brevity, we chose the proportional fairness function $\Sff = \sum_k \log(\bar{\Br}[k])$.

\section{Analytical Solutions}
\label{section:optimal_solution}

\subsection{Codeword Creation}
\label{section:codeowrd_creation}
Assume activation pattern $\Bp_j$ is given and helper $h_i$ is active in $\Bp_j$. Let us denote the set of users that can be served by $h_i$ as $\CU_i$; i.e., $\CU_i$ includes users within distance $r_{\mathrm{trans}}$ of $h_i$ but out of distance $r_{\mathrm{inter}}$ of every other active helper. Also, let us use $\CU_i^l$ to represent the subset of users of $\CU_i$ all assigned to the cache profile $l \in L$, i.e., for every $u_k \in \CU_i^l$ we have $\SfL(u_k) = l$. We use $\Sfl(i)$ to denote the number of $\CU_i^l$ sets for which $|\CU_i^l| = 0$.
The first step in codeword creation is to build a \emph{feasible} subset $\CV_i$ of $\CU_i$, where every user in $\CV_i$ is assigned to a different profile. Clearly, $|\CV_i| \le L-\Sfl(i)$, and the number of possible ways to build $\CV_i$ is $\prod_{l \in [L]} {(|\CU_i^l|+1)} - 1$, where addition with one (inside parentheses) is to account for the case no user is selected from $\CU_i^l$ and the subtraction of one is to exclude the empty set.

The users in a feasible set $\CV_i$ can be served by a number of codewords transmitted through helper $h_i$. To find these codewords, we build the \emph{support} set of $\CV_i$ as $\CL(\CV_i) = \{\SfL(u_k), \forall u_k \in \CV_i \}$, and then create its \emph{extended} set $\hat{\CV}_i$ by adding $L - |\CV_i|$ \emph{phantom} users $u_{l}^*$,\footnote{Phantom users are imaginary users added only to help with the formal definition of the delivery process. The same concept is used in~\cite{salehi2020lowcomplexity}.} $\SfL(u_l^*) = l$, for every $l \in [L] \backslash \CL(\CV_i)$. Next, a \emph{preliminary} codeword $\hat{X}(\hat{\CT}_i)$ can be built for every subset $\hat{\CT}_i$ of $\hat{\CV}_i$ with size $|\hat{\CT}_i| = t+1$ as
\begin{equation*}
    \hat{X}(\hat{\CT}_i) = \bigoplus_{u_{k} \in {\hat{\CT}_i}} W_{d_{k}, \CL(\hat{\CT}_i) \backslash \{\SfL(u_k)\} } ,
\end{equation*}
where $\CL({\hat{\CT}_i}) = \{\SfL(u_k), \forall u_k \in \hat{\CT}_i \}$ and $d_k$ is the index of the video chunk requested by $u_k$. Finally, the codeword $X(\CT_i)$ is built from $\hat{X}(\hat{\CT}_i)$ by removing the effect of phantom users ($\CT_i$ is the set resulting by removing phantom users from $\hat{\CT}_i$).
The codeword creation process clarifies exactly which codewords are generated for any given feasible set $\CV_i$.

\begin{exmp}
\label{exmp:codeword_creation}
    Consider the network in Example~\ref{exmp:placement} (visualized in Fig.~\ref{fig:network_model}). In Example~\ref{exmp:simple_codeword}, we provided transmitted codewords for this network if only $h_1$ is active. Now, let us consider $\Bp_j = [0,1]$. 
    For active helper $h_2$, we have $\CU_2^1 = \{u_4\}$, $\CU_2^2 = \{u_3,u_5\}$, and $\CU_2^3 = \varnothing$. We have five options to choose $\CV_i$; let us assume $\CV_2 = \{u_3,u_4\}$, and hence, $\CL(\CV_2) = \{1,2\}$ and $\hat{\CV}_2 = \{u_3,u_4,u_3^*\}$. We have three subsets $\hat{\CT}_2$ of $\hat{\CV}_i$ with size $t+1$, resulting in preliminary codewords
    \begin{equation*}
        \begin{aligned}
            \hat{X}(\{u_3,u_4\}) &= W_{d_3,\{1\}} \oplus W_{d_4,\{2\}}, \\
            \hat{X}(\{u_3,u_3^*\}) &= W_{d_3,\{3\}} \oplus W_{d_{3^*},\{2\}}, \\
            \hat{X}(\{u_4,u_3^*\}) &= W_{d_4,\{3\}} \oplus W_{d_{3^*},\{1\}}. \\
        \end{aligned}
    \end{equation*}
    After removing the effect of phantom user $u_3^*$, final codewords are $X(\{u_3,u_4\}) = W_{d_3,\{1\}} \oplus W_{d_4,\{2\}}$, $X(\{u_3\}) = W_{d_3,\{3\}}$, and  $X(\{u_4\}) = W_{d_4,\{3\}}$. Using the request pattern in Example~\ref{exmp:simple_codeword}, this means we must transmit $C_1 \oplus D_2$, $C_3$, and $D_3$.
\end{exmp}

\subsection{Rate Calculation}
Let us consider an activation pattern $\Bp_j$ and an active helper $h_i$ in this pattern. It can be easily verified that after the codeword creation process of Section~\ref{section:codeowrd_creation}, every user $u_k$ in the feasible set $\CV_i$ can successfully decode their requested chunk after $ \Sfn(\CV_i) = {\binom{L}{t+1} - \binom{L-|\CV_i|}{t+1}}$ codeword transmissions. Hence, the rate of every user $u_k \in \CV_i$ can be simply calculated as $r(\CV_i) = {1}/{\Sfn(\CV_i)}$. 
Now, to find the instantaneous rate vector $\Br(j,s)$, we need to 1) for every active helper $h_i$, choose the feasible set $\CV_i$, 2) for every $u_k \in \CV_i$, set the element $k$ of the vector $\Br(j,s)$ as $r(\CV_i)$, and 3) fill other elements of $\Br(j,s)$ as zero. Note that $r(\CV_i)$ is calculated according to the fact that $\binom{L-|\CV_i|}{t+1}$  preliminary codewords $\hat{X}(\hat{\CT}_i)$ are ignored as they include only phantom users.

\begin{exmp}
\label{exmp:rate_vector}
    Consider the same network and activation pattern as in Example~\ref{exmp:codeword_creation}. For active helper $h_2$, if $\CV_2 = \{u_3,u_4\}$, we get the three codewords shown in Example~\ref{exmp:codeword_creation}, and the rate vector is $[0,0,\frac{1}{3},\frac{1}{3},0]$. Similarly, it can be seen that for $\CV_2 = \{u_3\}$, we get two codewords $X(\{u_3\}) = W_{d_3,\{1\}}$ and $X(\{u_3\}) = W_{d_3,\{3\}}$, and the instantaneous rate vector is $[0,0,\frac{1}{2},0,0]$.
    
\end{exmp}

We call an instantaneous rate vector $\Br(j_1,s_1)$ to be \emph{dominated} by $\Br(j_2,s_2)$ if every element of $\Br(j_1,s_1)$ is smaller than or equal to the corresponding element of $\Br(j_2,s_2)$, and at least one element is strictly smaller. We call a rate vector \emph{maximal} if it is not dominated by any other rate vector.

Larger feasible sets $\CV_i$ result in more users being served and enhance multicasting. As a result, we restrict ourselves to maximal rate vectors that also maximize the coded caching gain for each helper. These vectors correspond to sets $\CV_i$ with the maximum possible length, i.e., $|\CV_i| = L-\Sfl(i)$.\footnote{The same fairness optimality may or may not be achieved after restricting the convex hull to the maximal rate vectors that also maximize the CC gain, depending on the network topology and the specific fairness function.} With this choice, the rates of users in $\CV_i$ become $r_{\Sfk(i)} = {1}/({\binom{L}{t+1}-\binom{\Sfl(i)}{t+1}})$. It can also be easily verified that with this restriction, the number of scheduling decisions per activation pattern $\Bp_j$ is upper bounded by
\begin{equation}
\label{eq:activation_pattern_count_maximal}
    \Sfc(\Bp_j) \leq \prod_{
    {i \in [H],
    \Bp_j[i] = 1}} \prod_{l \in [L]} {\max(|\CU_i^l|,1)} ,
\end{equation}
where the inequality results from the rate vectors that maximize the coded caching gain for each active helper but are not necessarily maximal. 

\subsection{The Optimization Problem}
\label{section:analytical_solution}
As mentioned in Section~\ref{section:sys_model}, in this paper, we aim to optimize the proportional fairness function over $\mathrm{Conv}(\Br(j,s))$. So, we need to solve the following convex optimization problem:
\begin{equation}
\label{eq:optimization_problem_main}
    \begin{aligned}
        \max_{a(j,s)} \Sff &= \sum_{k \in [K]} \log(\bar{\Br}[k]) \\
        s.t. \qquad & \bar{\Br} = \sum_{j \in [2^H]} \sum_{s \in [\Sfc(\Bp_j)]} a(j,s) \Br(j,s), \\
        & a(j,s)\geq{0}, \quad \sum_{j \in [2^H]} \sum_{s \in [\Sfc(\Bp_j)]} a(j,s) = 1 . \\
    \end{aligned}
\end{equation}
The problem with this optimization problem is its large dimensions: there exist $2^H$ activation patterns and the number of decisions per activation pattern in the order given by~\eqref{eq:activation_pattern_count_maximal}. This number grows very large, even for moderate-sized networks. 



\begin{rem}
Solving~\eqref{eq:optimization_problem_main} is not the only way to achieve the optimality of the fairness function. It can be done algorithmically, as time-averaged rates of the virtual queue method (Lyapunov Drift Plus Penalty) according to the well-known approach in~\cite{georgiadis2006resource}. However, such a scheduler needs to 
collect the status of the virtual queues for each user and 
solve at each scheduling slot the weighted sum rate problem over rate vectors, which requires the enumeration of all possible rate vectors and essentially has the same complexity of solving~\eqref{eq:optimization_problem_main} directly. The dynamic DPP scheduling algorithm has the advantage of being ‘on-line,' i.e., if the network topology evolves with time (e.g., users move and join/leave the network), the algorithm seamlessly adapts to the new topology (reflected by changes in the set of maximal rate vectors).
\end{rem}

\subsection{The Super-user Solution}
\label{section:super-user}
The complexity of the optimization problem~\eqref{eq:optimization_problem_main} can be reduced significantly by restricting the network throughput region (i.e., reducing $\mathrm{Conv}(\Br(j,s))$). The idea is that, for every activation pattern $\Bp_j$, we reduce the number of scheduling decisions $\Sfc(\Bp_j)$ to one by grouping users being served by the same active helper and assigned to the same cache profile into a \emph{super-user}. Then, the rate of every super-user is shared equally among all the users forming that super-user, as in the codeword formation, users sharing the same cache profile are interchangeable (i.e., the same codeword can carry data for any of these users).
This affects the instantaneous rate vectors: for every active helper $h_i$ in pattern $\Bp_j$, if $u_k \in \CU_i$, we fill the element $k$ of $\hat{\Br}(j,s)$ with ${r_{\Sfk(i)}}/{|\CU_i^{\SfL(u_k)}|}$.
The super-user approach imposes some ``local’’ fairness among the users sharing the same cache profile and served by a common active helper. However, it generally restricts the convex hull, making the throughput region smaller. As a result, the optimum of the fairness function could deteriorate compared with the solution of the original problem. 
%
%


\begin{exmp}
    Consider the same network and activation pattern as in Example~\ref{exmp:codeword_creation}. 
    Following the discussions in Example~\ref{exmp:rate_vector}, we have $r_{\Sfk(2)} = \frac{1}{3}$. As a result, the instantaneous rate vector $\hat{\Br}(j,s)$ for the super-user case is given as $[0,0,\frac{1}{6},\frac{1}{3},\frac{1}{6}]$.
\end{exmp}

\section{The ``Random Greedy Association" Algorithm}
\label{section:kagan_solution}
We propose a novel scheduling algorithm, called \emph{Random Greedy Association (RGA)} to find a sub-optimal solution to the optimization problem~\eqref{eq:optimization_problem_main}. This algorithm has low computational overhead and is applicable to large-scale networks with multiple helpers. Moreover, due to its underlying scheduling structure, it is applicable to dynamic network conditions where the users can move (and join/leave the network) freely or start/stop video streaming at any time.
The pseudo-code of the proposed RGA solver is provided in Algorithm~\ref{alg:main_body_inline}.
The description of new notations and their possible initial values (assigned with the $\textsc{Initialize}$ step) are as follows:
1) $\Bh_{\mathrm{on}}$ and $\Bh_{\mathrm{off}}$ include active and non-active helpers, respectively. $\Bh_{\mathrm{new}}$ also includes active helpers, but only the ones activated recently. Initially, all helpers are inactive; 
2) $\Sfv(u_k)$ denotes the number of chunks served to user $u_k$ and is initialized to zero. $v_{\mathrm{limit}}$ is a control parameter: the algorithm stops when $\Sfv(u_k) \ge v_{\mathrm{limit}}$ ,$\forall k \in [K]$;
3) $\Bd $ denotes the \emph{degree vector}. Defining the degree of user $u_k$, denoted as $\Sfd(u_k)$, as the number of helpers $h_i$ for which $|\Sfx(h_i)-\Sfx(u_k)| \le r_{\mathrm{inter}}$, $\Bd[s]$ corresponds to $s$-th smallest degree $\Sfd(u_k)$ for all $k\in{[K]}$; 
4) $\Bm_i$, $i \in [H]$, is a vector of size $L$, initialized to zero. $\Bm_i[l] = u_k$ only if helper $h_i$ is serving $u_k$ and $\SfL(u_k) = l$. Otherwise, it is zero; 
5) Auxiliary variables $\Sft(h_i)$ represent the amount of time helper $h_i$ requires to serve users in $\Bm_i$. Also, $T$ is the total delivery time, initially set to zero; and 
6) Auxiliary vector $\Bu_{\mathrm{cnd}}$, initially empty, is a placeholder for candid users to be served by the algorithm.

We have also used two auxiliary functions: $\textsc{Shuffle}(\Bv)$ randomly shuffles the order of data in $\Bv$ (used repeatedly during the algorithm runtime to avoid stalling), and $\textsc{Connect}(h_i,u_k)$ returns one only if $u_k$ is within radius $r_{\mathrm{trans}}$ of $h_i$ and outside of radius $r_{\mathrm{inter}}$ of every active helper in the network (other than --possibly-- $h_i$ itself), and zero otherwise. Finally, we have used a special notation in line 30 of Algorithm~\ref{alg:main_body_inline}: $[\Bm]$ denotes the set of all users $u_k$ that are receiving data from some helper.

As a brief explanation, at every iteration, the algorithm first checks if inactive helpers can be made active without causing interference to the active ones ($\textsc{AssignHelpers}$), and if that's possible, tries to maximize the number of users assigned to such recently activated helpers ($\textsc{AssignUsers}$). In both checks, users are filtered based on their degree values $\Sfd(\cdot)$, and it is ensured that every active helper serves at most one user assigned to every cache profile. A new iteration starts after at least one active helper finishes delivering video chunks, and it is assumed that the helpers that do not finish delivery remain active in the new iteration.\footnote{This is an important practical consideration as without that, we may need to further split subpackets into arbitrary smaller parts.} The algorithm runs as long as a minimum number of chunks are delivered to every user.

\begin{algorithm}[h]
\small
\caption{Random Greedy Association}
\label{alg:main_body_inline}
\begin{algorithmic}[1]
\State $\textsc{Initialize}$
\While{$\min_{k\in [K]} \Sfv(u_k) < v_{\mathrm{limit}}$}
\State $\textsc{Shuffle}(\Bh_{\mathrm{off}})$
\State $\textsc{Shuffle}(\Bd)$
\ForAll{$\hat{d} \in \Bd$}
\State $\textsc{AssignUsers}$
\State $\textsc{AssignHelpers}$
\EndFor
\ForAll{$h_i \in \Bh_{\mathrm{new}}$} $\Sft(h_i) \gets {1}/{r_{\Sfk(i)}}$ \EndFor
\State $\Bh_{\mathrm{new}} \gets [\;\;]$
\State $t_{\min} \gets \min_{h_i\in{\Bh_{\mathrm{on}}}} \Sft(h_i)$
\ForAll{$h_i\in{\Bh_{\mathrm{on}}}$}
  \State $\Sft(h_i) \gets \Sft(h_i) - t_{\min} $
  \If{$\Sft(h_i)=0$}
   \State Move $h_i$ from $\Bh_{\mathrm{on}}$ to $\Bh_{\mathrm{off}}$
    \ForAll{$l \in [L]$, $\Bm_i[l] \neq 0$}
            $\Sfv(\Bm_i[l]) \gets \Sfv(\Bm_i[l])+1$  
    \EndFor
    \State $\Bm_i \gets [0,0,\cdots,0]$
  \EndIf
\EndFor
\State $T \gets T+t_{\min}$
\EndWhile
\ForAll{$k\in [K]$}
  $r_k \gets \Sfv(u_k)/T$
\EndFor
\hrulefill
\Procedure{AssignUsers}{}
  \ForAll{$h_i \in \Bh_{\mathrm{new}}$}
   \ForAll{$k \in [K]$}
    \If{$\Sfd(u_k)=\hat{d}$, $\textsc{Connect}(h_i,u_k) = 1$}
     \State Add $u_k$ to $\Bu_{\mathrm{cnd}}$ 
    \EndIf
   \EndFor
   \State $\textsc{Shuffle}(\Bu_{\mathrm{cnd}})$
   \ForAll{$u_k \in \Bu_{\mathrm{cnd}}$}
     \If{$\Bm_i[\SfL(u_k)] = 0$}
       $\Bm_i[\SfL(u_k)] \gets u_k$
     \EndIf
   \EndFor
   \State $\Bu_{\mathrm{cnd}} \gets [\;\;]$  
  \EndFor
\EndProcedure
\hrulefill
\Procedure{AssignHelpers}{}
 \ForAll{$h_i \in \Bh_{\mathrm{off}}$}
  \If{$\nexists{u_k}\in{[\Bm]}$ s.t. $ |\Sfx(h_i)-\Sfx(u_k)| \le r_{\mathrm{inter}}$}
   \ForAll{$k \in [K]$}
    \If{$\Sfd(u_k) = \hat{d}$, $\textsc{Connect}(h_i,u_k) = 1$}
     \State Add $u_k$ to $\Bu_{\mathrm{cnd}}$    
    \EndIf
   \EndFor
   \State $\textsc{Shuffle}(\Bu_{\mathrm{cnd}})$ 
   \ForAll{$u_k \in \Bu_{\mathrm{cnd}}$}
     \If{$\Bm_i[\SfL(u_k)] = 0$}
       $\Bm_i[\SfL(u_k)] \gets u_k$
     \EndIf
   \EndFor
    \State $\Bu_{\mathrm{cnd}} \gets [\;\;]$   
  \EndIf
  \If{$\Bm_i$ is not empty}
   \State Remove $h_i$ from $\Bh_{\mathrm{off}}$ and add it to $\Bh_{\mathrm{on}}$, $\Bh_{\mathrm{new}}$
  \EndIf
 \EndFor
 \EndProcedure
\end{algorithmic}
\end{algorithm}

\section{Numerical Results}
\label{section:sim_results}




We use numerical results to compare the performance of various solutions. We assume $H$ helpers are located at the center of hexagons on a limited hexagonal grid. Every hexagon has a radius of~$1$, and we have $r_{\mathrm{trans}} = 1$ and $r_{\mathrm{inter}}  =1.2$. The users are placed according to a homogeneous Poisson Point Process in the transmission area of helpers. The average number of users per helper, shown by $U$, is determined by the density of the Poisson process. Every user is randomly assigned to a cache profile $l \in [L]$.

\begin{figure}[t]
    \centering
    \resizebox{0.75\columnwidth}{!}{%
    
    \begin{tikzpicture}

    \begin{axis}
    [
    axis lines = left,
    xlabel = \smaller {Rate=Delivered chunks per time},
    ylabel = \smaller {CDF},
    ylabel near ticks,
    legend pos = south east,
    ticklabel style={font=\smaller},
    grid=both,
    major grid style={line width=.2pt,draw=gray!30},
    ]
    
    
    
    \addplot[black]
    table[y=Opt-L1-Y,x=Opt-L1-X]{Figs/CDF_data_L.tex};
    \addlegendentry{\smaller $L=1$ \tiny (no CC)}
    \addplot[blue]
    table[y=Opt-L5-Y,x=Opt-L5-X]{Figs/CDF_data_L.tex};
    \addlegendentry{\smaller $L=5$}
    \addplot[green]
    table[y=Opt-L10-Y,x=Opt-L10-X]{Figs/CDF_data_L.tex};
    \addlegendentry{\smaller $L=10$}
    \addplot[red]
    table[y=Opt-L15-Y,x=Opt-L15-X]{Figs/CDF_data_L.tex};
    \addlegendentry{\smaller $L=15$}
    \addplot[gray]
    table[y=Opt-L20-Y,x=Opt-L20-X]{Figs/CDF_data_L.tex};
    \addlegendentry{\smaller $L=20$}

    \end{axis}

    \end{tikzpicture}
    }
    \caption{CDF for changing $L$, $U=10$, analytical solution, $\frac{M}{N} = 0.2$, Time is calculated assuming a full video chunk can be transmitted in one second.}
    \label{fig:cdf_l}
\end{figure}
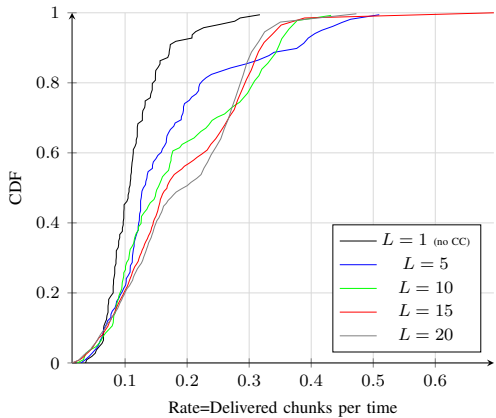

\begin{figure}[t]
    \centering
    \resizebox{0.75\columnwidth}{!}{%
    
    \begin{tikzpicture}

    \begin{axis}
    [
    axis lines = left,
    xlabel = \smaller {Rate= Delivered chunks per unit time of a codeword},
    ylabel = \smaller {CDF},
    ylabel near ticks,
    xticklabel style={
            /pgf/number format/fixed,
            /pgf/number format/precision=2
        },
        scaled x ticks=false,
    legend pos = south east,
    ticklabel style={font=\smaller},
    grid=both,
    major grid style={line width=.2pt,draw=gray!30},
    ]
    
    \addplot[black]
    table[y=Opt-U10-Y,x=Opt-U10-X]{Figs/CDF_data_U.tex};
    \addlegendentry{\smaller $U=10$, Analytical}
    
    \addplot[gray]
    table[y=Super-U10-Y,x=Super-U10-X]{Figs/CDF_data_U.tex};
    \addlegendentry{\smaller $U=10$, Super-user}
    
    \addplot[blue]
    table[y=Heur-U10-Y,x=Heur-U10-X]{Figs/CDF_data_U.tex};
    \addlegendentry{\smaller $U=10$, RGA}
    
    \addplot[blue,dashed]
    table[y=Reuse-U10-Y,x=Reuse-U10-X]{Figs/CDF_data_U.tex};
    \addlegendentry{\smaller $U=10$, Multi-round~\cite{mozhgan}}

    \addplot[green]
    table[y=Super-U20-Y,x=Super-U20-X]{Figs/CDF_data_U.tex};
    \addlegendentry{\smaller $U=20$, Super-user}
    
    \addplot[red]
    table[y=Heur-U20-Y,x=Heur-U20-X]{Figs/CDF_data_U.tex};
    \addlegendentry{\smaller $U=20$, RGA}

    \addplot[red,dashed]
    table[y=Reuse-U20-Y,x=Reuse-U20-X]{Figs/CDF_data_U.tex};
    \addlegendentry{\smaller $U=20$, Multi-round~\cite{mozhgan}}

    \end{axis}

    \end{tikzpicture}
    }
    \caption{CDF for changing $U$, $L=5$, $\frac{M}{N}$ = 0.2.}
    \label{fig:cdf_u}
\end{figure}
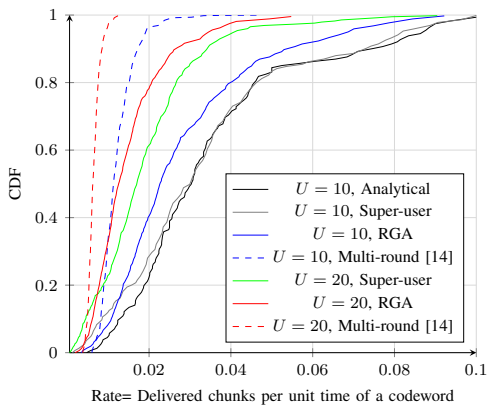

In Tables~\ref{tab:utiliy_u} and~\ref{tab:utiliy_l}, we have compared the utility function value, versus $U$ and $L$, of the analytical (Section~\ref{section:analytical_solution}), super-user (Section~\ref{section:super-user}), and RGA (Section~\ref{section:kagan_solution}) solutions, as well as multi-round delivery solution of~\cite{mozhgan} with a frequency reuse factor of three. 
Notice that $L=1$ corresponds to the uncoded caching scheme, i.e. we have just unicast opportunities.  For the analytical solution, we have limited $\mathrm{Conv}(\Br(j,s))$ to maximal rate vectors that also maximize the coded caching gain, as the more general case is intractable even for small networks due to its complexity (even with the restriction, we cannot simulate large networks as is visible in the table). 
We can see that:

\noindent $\bullet$ Both RGA and super-user perform well; for small networks, their performance is quite close to the analytical solution. Moreover, in all cases, RGA performs at most $10\%$ worse than the super-user solution;

\noindent $\bullet$ RGA, analytical and super-user solutions significantly outperform the baseline multi-round delivery solution;

\noindent $\bullet$ System performance gets worse as $U$ is increased (even if we normalize utility values with $U$, the same conclusion is persistent). This is expected, as with more users, the portion of resources allocated to each user becomes more limited;

\noindent $\bullet$ System performance improves as $L$ is increased. This is due to the increased multicasting opportunities as $L$ grows; and

\noindent $\bullet$ As $L$ keeps growing, the increase in the utility function slows down irrespective of the selected solver. This is because $U$ is fixed, and hence, the multicasting opportunities are limited even if $L$ gets very large. 


In Figure~\ref{fig:cdf_l}, we have extended the comparisons by plotting the CDF of the achievable rate (in terms of delivered chunks per unit time) of all users, for the analytical solver and for $L \in \{1,5,10,15,20\}$ (in comparisons, we take into account the size of the codewords and normalize accordingly).   
Similarly, in Figure~\ref{fig:cdf_u}, we have plotted the CDF of the achievable rates (in terms of the number of delivered chunks) of all users, for different solvers and for $U \in \{10,20\}$.
These CDF plots essentially confirm the observations made by Tables~\ref{tab:utiliy_u} and~\ref{tab:utiliy_l}: heuristic algorithms perform well, and analytical and heuristic methods outperform the baseline solution by a good margin.

\begin{table}[t]
    \centering
    \vspace{3pt}
    \begin{tabular}{c||c|c|c|c|c|c}
         $U$ & 5 & 10 & 15 & 20 & 25 & 30  \\
         \hline
         \hline
         Analytical & 59 & 138 & & & &  \\
         \hline
         Super-user & 60 & 142 & 230 & 338 & 446 & 548 \\
         \hline
         RGA & 63 & 147 & 238 & 349  & 462  & 568  \\
         \hline
         Multi-round~\cite{mozhgan} & 80 & 171 & 279 & 404 & 530 & 647
    \end{tabular}
    \caption{$\mathrm{round}(|f|)$ vs average users per helper $U$, $L=5$.}
    \label{tab:utiliy_u}
\end{table}

\begin{table}[t]
    \centering
    \begin{tabular}{c||c|c|c|c|c}
         $L$ & 1 & 5 & 10 & 15 & 20 \\
         \hline
         \hline
         Analytical &  86 & 75 & 72 & 70 & 69 \\
         \hline
         Super-user & 95 & 78 & 74 & 72 & 71 \\
         \hline
         RGA & 98 & 84 & 81 & 79 & 78 \\
         \hline
         Multi-round~\cite{mozhgan} & 127 & 112 & 107 & 104 & 102
    \end{tabular}
    \caption{$\mathrm{round}(|f|)$ vs cache profile count $L$, $U=10$.}
    \label{tab:utiliy_l}
\end{table}






\section{Conclusion}
Considering coded caching techniques for on-demand video streaming over WLANs where multiple users are served simultaneously by multiple spatially distributed APs, we formulated the region of achievable long-term average delivery rate (defined as the number of video chunks delivered per unit of time) and studied the per-user rate distribution under proportional fairness scheduling. We also developed reduced complexity scheduling strategies and compared them with standard state-of-the-art techniques such as conventional (uncoded) caching and collision avoidance by allocating APs on different sub-channels (i.e., frequency reuse). Simulation results confirmed the performance of the proposed schemes.

\bibliographystyle{IEEEtran}
\bibliography{references,kagan_references}


\end{document}